\documentclass[12pt]{amsart}
\usepackage{amssymb, latexsym}

\newcommand{\R}{{\mathbb R}}
\newcommand{\C}{{\mathbb C}}
\newcommand{\T}{{\mathbb T}}

\newcommand{\J}{{\mathbb J}}

\newcommand{\loc}{ {\mbox{\scriptsize{loc}}} }

\newcommand\half{\frac{1}{2}}
\renewcommand{\div}{\mbox{div}\;}
\renewcommand{\det}{\mbox{det}\;}

\newcommand\ue{{u^{\epsilon}}} 
       
\newcommand\e{\epsilon}
\newcommand\en{\epsilon_n}
\newcommand\phie{\phi^\e}
\newcommand\psie{{\psi^\e}}

\newcommand\Ttilde{\widetilde{T}}
\newcommand\uen{{u^{\epsilon_n}}}

\newtheorem{theorem}{Theorem}
\newtheorem{proposition}{Proposition}
\newtheorem{lemma}{Lemma}
\theoremstyle{definition}

\theoremstyle{remark}
\newtheorem{remark}{Remark}
\numberwithin{equation}{section}

\begin{document}

\title{Vortex dynamics for the Ginzburg-Landau-Schr\"odinger equation}
\author{J.~E.~ Colliander}
\address{\hskip-\parindent J.~E.~ Colliander\\
Mathematical Sciences Research Institute, Berkeley, CA 94720}
\author{R.~L.~ Jerrard}
\address{\hskip-\parindent R.~L.~ Jerrard\\
University of Illinois, Urbana, IL 61801}
\begin{abstract}
 
The initial value problem for the Ginzburg-Landau-Schr\"odinger equation
is examined in the $\e \rightarrow 0$ limit under two main assumptions on the
initial data $\phie$. The first assumption is that $\phie$ exhibits 
$m$ distinct vortices of degree $\pm 1$; these are described as points 
of concentration of the Jacobian $[J\phie]$ of $\phie$. Second,
we assume energy bounds consistent with vortices at the points of 
concentration. Under these assumptions, we identify ``vortex structures''
in the $\e \rightarrow 0$ limit of $\phie$ and show that these structures
persist in the solution $\ue(t)$ of $GLS_\e$. We derive ordinary differential
equations which govern the motion of the vortices in the $\e \rightarrow 0$
limit. The limiting system of ordinary differential equations is a Hamitonian
flow governed by the renormalized energy of Bethuel, Brezis and H\'elein.
Our arguments rely on results about the structural stability of vortices 
which are proved in a separate paper.

\end{abstract}

\maketitle

\section{Introduction}

We investigate the $\e \rightarrow 0$ behavior of the initial value problem
\begin{equation}
\left\{ \matrix
i{u^\epsilon_t} - \Delta \ue + {\frac{1}{\epsilon^2}} {{\left( {{|\ue
|}^2} -1 \right)}} \ue = 0, & {u^\e}:{\T^2} \times [0,T) \longmapsto {\R^2} \\
\ue(x,0) = {\phi^\e}(x), & x \in {\T^2}
\endmatrix
\right.
\tag*{${{GLS}}_{\epsilon}$}
\end{equation}
exhibiting a finite number of vortices, under appropriate assumptions
on the associated Hamiltonian
\begin{equation}
{I^\e}[u] = {\int\limits_{\T^2}}  {E^\e} (u) dx;~~{E^\e} (u) = \half
{{|D\ue|}^2} + {\frac{1}{4{\e ^2}}} {{\left({{|\ue|}^2}-1\right)}^2}.
\end{equation}

The $\e \rightarrow 0$ asymptotics of minimizers of ${I^\e}[u]$
subject to Dirichlet boundary constraints were studied by Bethuel, Brezis and H\'elein in
\cite{BetBreHel94}. The corresponding parabolic evolution problem was
studied by Lin \cite{Lin94.2} and Jerrard and Soner \cite{JerSon95.1}. These works 
identify ``vortex structures'' emerging in the $\e \rightarrow 0$ limit whose 
location is governed by a renormalized energy. The renormalized energy is
obtained by removing the divergent part of ${I^\e}[u]$. This note and
the forthcoming paper \cite{ColJer97}
establish similar results for the
conservative evolution ${GLS}_\e$.

The Landau theory of second order phase transitions \cite{Landau41}
consists of expanding the energy in terms of a parameter which encodes
the ``order'' in the phase and then exploiting energy properties to
determine the evolution of the ``order parameter''. This theory was
applied by Ginzburg and Pitaevskii \cite{GinPit58} and Pitaevskii
\cite{Pit61} to argue that the order parameter describing superfluid
helium II evolves according to $GLS_\e$. In this context,
${I^\e}$ is the free energy and $\ue$ is the order parameter ``which
plays the role of `the effective wave function' of the superfluid part
of the liquid'' \cite{GinPit58}. The motion of $\ue$ under the $GLS_\e$
evolution conserves ${I^\e}[\ue]$. If we express
$\ue({x_1},{x_2}) = {\rho({x_1},{x_2})} {e^{i \theta({x_1},{x_2})}},$
with $\rho, \theta$ $\R$-valued, then $\rho^2$ represents the density of the
superfluid and $D \theta$ is the velocity of the superfluid. 
Gross \cite{Gro66} also derived ${GLS}_\e$
as the Schr\"odinger equation
for a wave function describing a system of interacting bosons. 
The equation ${GLS}_\e$ is often called the Gross-Pitaevskii equation
in the physics literature.

Because we are interested in studying the motion of vortices in the 
$\e \rightarrow 0$ limit, we choose initial data ${\phi^\e}: {\T^2} 
\longmapsto \C$ possessing vortices.
This heuristic notion is implemented by imposing two conditions.
First, we assume the measure $[J{\phi^\e}]= \det (D{\phi^\e})~ dx$
is close to a weighted sum of Dirac masses. More precisely, we assume
\begin{equation}
 [J{\phi^\e}] \rightharpoonup  \pi {\sum\limits_{i=1}^m} {d_i} {\delta_{\alpha_i}}
\label{JtoDirac}
\end{equation}
where ${d_i} = {\pm 1}$ with ${\sum\limits_{i=1}^m} {d_i} = 0$, and
${\alpha_i}$ are distinct points in ${\T^2}$.

Second, we assume energy bounds. For example,
certain of our results
require that ${\phi^\e}$ satisfy
\begin{equation}
{I^\e}[{\phi^\e}] \leq \pi m \log {\frac{1}{\e}} + {\gamma_1}.
\label{Ieupper}
\end{equation}
for some $\gamma_1 \in {\R}$.
It follows from results that are cited and discussed
in Section 3 that
assuption \eqref{JtoDirac} implies the lower bound
\begin{equation}
{I^\e}[{\phi^\e}] \geq \pi m \log {\frac{1}{\e}} - {C},
\label{Ielower}
\end{equation}
where ${C}$ is independent of $\e$. 
Thus \eqref{Ieupper} asserts that the energy of $\phi^\e$
is $O(1)$, given the prescribed configuration of vortices.
Our strongest results need more stringent hypotheses,
which amount to assuming that the energy is $o(1)$, relative
to the prescribed configuration of vortices.

We briefly outline the contents of this paper. Since
we have not yet defined many of the terms we are using,
this description is necessarily a little impressionistic.

Section 2 introduces notation and records some
useful identities satisfied by solutions of ${GLS}_\e$.

Section 3 states and briefly discusses some results 
providing conditions under which the Jacobian of a function $u$
(which in our context may be interpreted as the vorticity)
and the Ginzburg-Landau energy density concentrate around a
collection of points. These results provide a key element in
our analysis. The proofs are long and technical and will
appear in \cite{ColJer97}. Section 3 also states some
refined estimates related to the renormalized energy
introduced by Bethuel, Brezis, and H\'elein \cite{BetBreHel94}.
These too are proven in \cite{ColJer97}.

In Section 4 we prove that solutions of $GLS_\e$ exhibit
Lipschitz paths along which the energy and the vorticity
concentrate. The main point is to control vortex mobility.
We also describe weak limits of the ``supercurrent''
associated with solutions.

Finally, in Section 5 we give a complete description
of limiting vortex dynamics, under somewhat stronger assumptions
on the initial data. We also sharpen our characterization
of the limiting behavior of solutions away from the vortices.

We ultimately show that, under appropriate
hypotheses, the limiting vortices solve
a system of ODEs, which is exactly that satisfied by
classical point vortices in an ideal two dimensional fluid.
This result was predicted as early as 1965 by Fetter
\cite{Fetter65}, based on physical arguments, and
more recently by Neu \cite{Neu90} and E \cite{E94},
who arrived at the same conclusion through matched asymptotics.

\section{Notation and identities}

We introduce some vector notation and the quantity $j(\phie)$.
We then define measures $[J \phie ]$ and $\mu^\e_{\phie}$ which will be used to locate the singularities of $\phie$. Then a norm is introduced that permits us to say the singularities of two functions $\psie$ and $\phie$ are close if the associated measur
es $[J \psie],~{\mu^\e_\psie}$ and $[J \phie ],~{\mu^\e_{\phie}}$ are close. Next, we derive some identities for $[J {\ue} ]$ and ${\mu_{\ue}^\e}$ assuming $\ue$ evolves according to ${GLS}_\e$. Finally, we introduce 
the renormalized energy $W$ and a canonical harmonic map $H$ with singularities of degrees ${\{{d_1}, \dots , {d_m} \}}$ located at the points ${\{{a_1}, \dots , {a_m} \}}$ mapping ${\T^2}\setminus{\{{a_1}, \dots , {a_m} \}} \longmapsto {S^1}$.

We introduce some notation for vector fields $u:{\T^2} \longmapsto
{\R^2}$ mapping $({x_1},{x_2}) \in {\T^2}$ to $({u^1}({x_1},{x_2}), {u^2}({x_1},{x_2})) \in {\R^2}$. The scalar product in ${\R^2}$ is denoted by ``$\cdot$'',
\begin{equation*}
u \cdot v = {u^1}{v^1} +{u^2}{v^2}.
\end{equation*}
Sometimes we will view $u \in {\R^2}$ as $u \in {\C}$ when we write $iu$ or ${e^{i \alpha}} u$. These expressions are interpreted in the obvious way, e.g.
\begin{equation*}
iu = -\J u;~~~~~\J= \left( \matrix 0 & 1 \\
                   -1 & 0  \endmatrix
    \right).
\end{equation*}

We define for $u,v \in {\R^2}$, 
\begin{equation*}
u \times v = {u^1}{v^2} - {u^2}{v^1} = {\J_{ij}} {u^i}{v^j},
\end{equation*}
\begin{equation*}
\nabla \times u = {\partial_{x_1}}{u^2} - {\partial_{x_2}}{u^1}.
\end{equation*}
Repeated indices are implicitly summed throughout this paper.
Note that $(iu)\cdot v= u \times v$ and $(iu) \cdot u  = 0$. For a scalar function $\phi$ we define
\begin{equation*}
\nabla \times \phi = ({\phi_{x_2}}, -{\phi_{x_1}}).
\end{equation*}

For a sufficiently differentiable $u:{\T^2} \longmapsto {\R^2}$, we define
\begin{equation*}
j(u) = (u \times {u_{x_1}}, u \times {u_{x_2}}).
\end{equation*}
We sometimes write $j(u) = (iu) \cdot Du . $ In the superfluid model, $j(\ue)$ is interpreted as a current. If we write $u= \rho {e^{i \theta}}$, for $\rho, \theta \in \R$, then $j(u) = {\rho^2} D\theta.$ In particular, if $|u| =1$, $j(u)$ is the phase gr
adient.

We define the signed Jacobian of $u$,
\begin{equation}
Ju = \det Du.
\end{equation}
The quantities $j(u)$ and $Ju$ are related by the identity
\begin{equation}
\nabla \times j(u) = 2 {u_{x_1}} \times {u_{x_2}} = 2 Ju,
\label{Jfromj}
\end{equation}
so the signed Jacobian can also be interpreted as the vorticity.
We write $[Ju]$ to denote the distributional signed Jacobian, defined if
$j(u) \in {L^1}$ via
\begin{equation}
\int \phi [Ju] = \half \int (\nabla \times \phi) \cdot j(u),~~~\phi \in {C^1_c}({\T^2}).
\label{[J]def}
\end{equation}
In particular, $[Ju]$ is well defined for $u\in {H^1}({\R^2})$. 

For a given $u \in {H^1}({\T^2}; {\R^2})$, define the measure
\begin{equation*}
{\mu_u^\e}(A) = {\frac{1}{|\log \e |}} {\int\limits_{{A}} }
{E^\e}(u) ~dx;~{E^\e}(u)= \half {{|Du|}^2} + {\frac{1}{4 {\e^2}}} {{({{|u|}^2} -1 )}^2} ,
\end{equation*}
for subsets $A \subset {\T^2}$. The renormalization factor ${\frac{1}{|\log \e |}}$ appears naturally upon considering $u = {\frac{x}{|x|}}$ smoothly cutoff in a ball of radius $\e$ centered at $x=0$.

We let $M({\T^2})$ denote the dual of ${C}({\T^2})$, i.e. the space of signed Radon measures on ${\T^2}$. Similarly, we let ${M^1}({\T^2})$ denote the dual of ${C^1}({\T^2})$, equipped with the dual norm. We also define a seminorm
\begin{equation*}
{{\| \mu \|}_{{{\widehat{M}}^1}({\T^2})}} = \sup \left\{ \int \phi ~d\mu ~:~ {{\| D \phi \|}_{L^\infty}} \leq 1,~ \int \phi = 0  \right\}.
\end{equation*}
If $\mu({\T^2}) = 0$, we can compute ${{\|\mu\|}_{{{M^1}({\T^2})}}}$ by testing $\mu$ against functions $\phi$ such that $\int \phi~dx = 0$. In this case it follows that
\begin{equation*}
{c^{-1}} {{\| \mu \|}_{{{\widehat{M}}^1}({\T^2})}} \leq 
{{\| \mu \|}_{{{{M}}^1}({\T^2})}} \leq {{\| \mu \|}_{{{\widehat{M}}^1}({\T^2})}}.
\end{equation*}
This is a consequence of the fact that ${{\| \phi \|}_{{C^1}({\T^2})}} \leq C {{\|D \phi \|}_{{L^\infty}({\T^2})}} $ whenever $\int \phi = 0$.

In particular, if $\mu$ has the form
\begin{equation}
\mu = {\sum\limits_{i=1}^n} {\delta_{\xi_i}} -  {\sum\limits_{i=1}^n} {\delta_{\eta_i}}
\label{muform}
\end{equation}
for some points ${\xi_1}, \dots, {\xi_n}, {\eta_1}, \dots, {\eta_n} \in {\T^2}$, not necessarily distinct, then Brezis, Coron and Lieb \cite{BreCorLie86} show that
\begin{equation}
{{\| \mu \|}_{{{\widehat{M}}^1}({\T^2})}} = {\min\limits_{\pi \in {S_n}}} {\sum\limits_{i=1}^n} |{\xi_i} - {{\eta_{\pi(i)}}}|
\label{bcl}
\end{equation}
where $S_n$ is the symmetric group of permutations of $n$ objects. For measures of the form \eqref{muform} we thus have
\begin{equation*}
{c^{-1}} {\min\limits_{\pi \in {S_n}}} {\sum\limits_{i=1}^n} |{\xi_i} - {{\eta_{\pi(i)}}}| \leq {{\| \mu \|}_{{M^1}({\T^2})}} \leq {\min\limits_{\pi \in {S_n}}} {\sum\limits_{i=1}^n} |{\xi_i} - {{\eta_{\pi(i)}}}| .
\end{equation*}
This estimate permits us to prove a fact illustrating the usefulness of the $M^1$ norm.
\begin{proposition}
Suppose that for every $t \in [0,T),~{\mu_t}$ is a measure of the form
${\sum\limits_{i=1}^n} {\delta_{{a_i}(t)}}$, for certain points ${{a_1}(t)}, \dots, {{a_n}(t)}.$ The ${\mu_{(\cdot)}}$ is a countiuous (resp. Lipschitz) function from $[0,T)$ into $M^1$ if and only if the points ${a_i}(t)$ can be labelled in such a way th
at ${a_i}(\cdot)$ is continuous (resp. Lipschitz) for each $i$.
\end{proposition}
\begin{proof}
For any $s,t$, the measure ${\mu_t}-{\mu_s}$ has integral zero and so satisfies \begin{equation*}
{{\| {\mu_t}-{\mu_s} \|}_{{M^1}({\T^2})}} \thicksim 
{\min\limits_{\pi \in {S_n}}} {\sum\limits_i} |{a_i}(t)-{a_{\pi(i)}}(t)|.
\end{equation*}
The lemma follows immediately.
\end{proof}

\begin{remark}
The ${\widehat{M}}^1$ seminorm can be interpreted as the minimum cost in a Monge-Kantorovich mass transfer problem. Indeed, when $\int d\mu=0$ as above, $\mu$ can be written in the form $\mu={\nu_1}-{\nu_2}$, where ${\nu_1}$ and ${\nu_2}$ are positive, mu
tually singular measures and ${\nu_1}({\T^2}) = {\nu_2}({\T^2})$. Then ${{\| \mu \|}_{{\widehat{M}}^1}}$ is precisely the minimum cost of ``transporting'' ${\nu_1}$ to ${\nu_2}$, subject to an appropriate cost functional. See for example \cite{EvaGan96} f
or a more precise statement and more details.
\label{monge}
\end{remark}
\begin{remark}
Any reasonable weak norm on measures would be equally suitable for our purposes. The $M^1$ norm is a convenient choice, but is certainly not the only possible choice.
\label{reasonable}
\end{remark}

We employ $O$ and $o$ notation in some of the analysis below. We write, for example, ${O_{a,b,c}}(1)$ to indicate a quantity is $O(1)$ with respect to the interesting limit $\e \rightarrow 0$, with the implicit constant depending only upon the parameters 
$a,b,c.$

We assume in this section, for fixed $\e > 0$, that $u$ solves
\begin{equation*}
i{u_t} - \Delta u + \frac{1}{\e^2} \left({{|u|}^2} -1 \right) u = 0
\end{equation*}
and derive evolution equations for certain functions of $u$ and its
derivatives. These are mostly well known. We formally calculate the evolution equation for ${E^\e}(u)= \half {{|Du|}^2} + \frac{1}{4 {\e^2}} {{\left( {{|u|}^2}-1 \right)}^2}. $
\begin{eqnarray*}
\frac{d}{dt} {E^\e}(u) & = &{u_{x_i}}\cdot{u_{{x_i}t}} + \frac{1}{\e^2} ({{|u|}^2}-1) u \cdot {u_t}\\
& = &{{({u_{x_i}}\cdot{u_t})}_{x_i}} - ({u_{{x_i}{x_i}}} - \frac{1}{\e^2} ({{|u|}^2}-1)) \cdot {u_t}
\end{eqnarray*} 
so that
\begin{equation}
\frac{d}{dt} {E^\e}(u)= {{({u_{x_i}}\cdot{u_t})}_{x_i}}
\label{Eev}
\end{equation}
using the equation and $(i{u_t})\cdot {u_t}=0$.

Similarly, we calculate
\begin{equation*}
\frac{d}{dt} \half {{|u|}^2} = u \cdot {u_t} = -(iu)\cdot \Delta u 
\end{equation*}
to obtain
\begin{equation}
\frac{d}{dt} \half {{|u|}^2} = -{{((iu)\cdot {u_{x_i}})}_{x_i}}= -\div ~ j(u).
\label{L2ev}
\end{equation}

Next, we consider the evolution of a component of $j(u)$.
\begin{eqnarray*}
\frac{d}{dt}{j^k}(u)& = &\frac{d}{dt}(iu)\cdot {{u_{x_k}}} = (i{u_t})\cdot {u_{x_k}} + (iu)\cdot {u_{t{x_k}}} \\
& = & (i{u_t})\cdot {u_{x_k}} - u \cdot {{(i{u_{t{x_k}}})}} \\
& = & {u_{{x_j}{x_j}}}\cdot {u_{x_k}} - u\cdot {u_{{x_j}{x_j}{x_k}}} + \frac{1}{\e^2} {{|u|}^2} {{|u|}^2_{x_k}} \\
& = & {{({u_{x_k}}\cdot{u_{x_j}} - u\cdot{u_{{x_k}{x_j}}})}_{x_j}} + \frac{1}{\e^2} {{|u|}^4_{x_k}}.
\end{eqnarray*}
We rewrite as
\begin{equation}
\frac{d}{dt}{j^k}(u) = 2 {{({u_{x_k}}\cdot{u_{x_j}})}_{x_j}}
 - {{({{|{u_{x_j}}|}^2} - u\cdot{u_{{x_j}{x_j}}})}_{x_k}} + \frac{1}{2{\e^2}} {{|u|}^4_{x_k}}.
\label{jev}
\end{equation}
Note that this shows ${\int_{{\T^2}}} j(u) dx$ is time-independent.

We calculate the evolution equation of $Ju$ using \eqref{Jfromj}
\begin{equation*}
\frac{d}{dt} Ju = \nabla \times {{(Du \cdot {u_{x_j}})}_{x_j}},
\end{equation*}
which we write out explicitly,
\begin{equation}
\frac{d}{dt} Ju = {{ {\J_{jk}} {{{\left({u_{x_k}} \cdot {u_{x_l}}
\right)}_{{x_l}{x_j}}}} }} .
\label{jacid}
\end{equation}

Finally, we express \eqref{jacid} in a weak form.
Multiply \eqref{jacid} by $\eta \in {C^\infty_c}({\T^2})$ and integrate to find
\begin{equation}
\frac{d}{dt} {\int_{\T^2}} \eta ~Ju ~dx = {\int_{\T^2}} {\eta_{{x_j}{x_l}}} {\J_{jk}}{u_{{x_k}}} \cdot {u_{{x_l}}}~dx.
\label{wjacid}
\end{equation}
This is the key identity in our analysis of vortex dynamics of $GLS_\e$.
\begin{remark} The preceding calculations apply equally well to
$\ue(\cdot) \in {H^1}({\T^2})$. For $\e > 0$ fixed, $GLS_\e$ is a
defocussing nonlinear Schr\"odinger equation. Bourgain has established
\cite{B1} global wellposedness for $GLS_\e$ below $H^1$. We validate
the preceding calculations for $\ue \in {H^1}({\T^2})$ as follows
using various aspects of Bourgain's result. By continuous dependence on
the data, a different solution ${\widetilde{\ue}}(t)$ is close to $\ue(t)$ in
${H^1}({\T^2})$ provided the corresponding initial data
${\widetilde{\phie}}$ and $\phie$ are close in ${H^1}({\T^2})$. Let
${\widetilde{\phie}}$ be a smooth approximator to $\phie$. The
preceding calculations apply to ${\widetilde{\ue}}$ since it remains
smooth for all time. The various identities above, in particular
\eqref{wjacid}, are then validated for $\ue \in {H^1}({\T^2})$ by considering
a sequence of smooth approximators.
\label{Honeev}
\end{remark}

We reformulate for the torus $\T^2$ some of the notions introduced by Bethuel, Brezis and H\'elein \cite{BetBreHel94} for the study of Dirichlet minimizers of ${I^\e} [u]$. 
Let $F$ solve
\begin{equation}
\Delta F = 2 \pi( \delta_0 - 1 )
\hspace{4em}\mbox{on }{\T}^2.
\label{Fund}
\end{equation}
with the normalization
\begin{equation}
\lim_{x\rightarrow 0} (F(x) -  \log |x|) = 0
\label{normalize}\end{equation}

For points $ a_1, ..., a_m \in {\T^2}$ (sometimes denoted by $(a)$)
and nonzero integers $d_1, ..., d_m$ satisfying
$\sum d_i = 0$, define the renormalized energy
\begin{equation}
W(a, d) := -  \pi \sum_{i \ne j} d_i d_j F(a_i - a_j).
\label{W.def}\end{equation}

%
%
Define 
\begin{equation}
\Phi =   \sum_{i=1}^m d_i F(x-{a_i})
\label{Phi}
\end{equation}
and notice that
\[
\Delta \Phi = 2 \pi {\sum_{i=1}^m} {d_i} {\delta_{a_i}},~~~~~\int \Phi = 0.
\]

The following proposition is proved in \cite{BetBreHel94} (Theorem 1.1., Step 2 of proof).

\begin{proposition}
There is a map $H \in C^\infty_\loc({ \T}^2 \setminus (a); S^1)
\cap W^{1,1}({\T}^2; S^1)$ satisfying
\begin{equation}
j(H) = -\nabla \times \Phi.
\label{Hdefined}\end{equation}
Moreover, if ${\widetilde{H}}$ is any other function satisfying \eqref{Hdefined}, then ${\widetilde{H}} = {e^{i \alpha}} H$ for some $\alpha \in \R$.
\end{proposition}

It follows from \eqref{Hdefined} that $H$ satisfies
\begin{equation}
\div j(H) \; = \; 0,
\label{chm1}\end{equation}
\begin{equation}
2 [JH] = \nabla\times j(H) = 2\pi \sum_{i=1}^m d_i \delta_{a_i}.
\label{chm2}\end{equation}
in the sense of distributions
and pointwise away from the singluarities
$(a)$.
The map $H$ also satisfies
\begin{equation}
\int_{ {\T}^2} j(H) dx = 0.
\label{chm3}\end{equation}

The map $H$ has singularities of degree $(d)$ located at the points $(a)$ as seen in \eqref{chm2}. The identity \eqref{chm1} reveals that $H$ is a harmonic map into $S^1$ since $j(H) = D \theta$ when we express $H={e^{i \theta}}$. Therefore, we refer to $
H$ as the {\em{canonical harmonic map}}\begin{footnote}{We have taken the name
``canonical harmonic map'' from the work of Bethuel, Brezis, and
H\'elein on the Dirichlet problem, where it is entirely appropriate. In
the periodic context it is clearly something of a misnomer, since
$H$ is not unique. Nonetheless, it seems easiest to
use the familiar terminology.}\end{footnote} with singularities of degree $(d)$ located at the points $(a)$.
We will sometimes highlight the dependence of $H$ upon $(d)$ and $(a)$ by writing $H(x ; a,d)$.

For a collection of points $a_1, ..., a_m \in \T^{2}$
define
\[
{\T}^2_\rho := {\T}^2 \setminus \cup_i B_\rho(a_i).
\]
We will be interested in the case
\begin{equation}
0 < \rho < \frac 14 \min_{i\ne j} |a_i - a_j|.
\label{small.rho}\end{equation}

A configuration of vortices $(a),(d)$ determines (up to a constant phase factor $e^{i \alpha}$) a canonical harmonic map $H(\cdot,a,d)$. The map $H$ has logarithmically divergent Dirichlet energy. The renormalized energy $W(a,d)$ describes to leading orde
r the finite part of the energy associated with the configuration $(a),(d)$. The following proposition, essentially proven in \cite{BetBreHel94} (Theorem I.7), makes these observations precise. 
\begin{proposition} Let $H = H(\cdot, a,d)$.
For $\rho$ satisfying \eqref{small.rho},
\[
\int_{\T^2_\rho} \frac 12 |D H|^2 dx = m \pi \log {\frac{1}{\rho}} + W(a,d) + O(\rho).
\]
\end{proposition}

%

We define, following Bethuel, Brezis, and H\'elein \cite{BetBreHel94}, the quantity 
\begin{equation}	
I(\e, \rho) = \min \left\{ \int_{B_\rho} { E}^\e(u) dx \; : \;
	u\in H^1(B_\rho), ~u(x) = \frac x{|x|}
	\mbox{ for }x\in \partial B_\rho \right\}.
\label{I.def}\end{equation}
It is shown in \cite{BetBreHel94} that
\begin{equation}
I(\e,\rho) = \pi \log (\frac \rho \e) + O(1)
\label{Ier.est}\end{equation}
as $\e\rightarrow 0$, for $\rho$ fixed.

An easy adaptation of the construction given in (\cite{BetBreHel94}, Lemma VIII.1) establishes

\begin{proposition}
Given any distinct $a_1, ... ,a_m \in \T^2$ and
$d_1, ... , d_m \in \{ \pm 1 \}$, there exists a 
family of functions $v^\e \in H^1(\T^2; \R^2)$
such that
\[
[J v^\e] \rightharpoonup {\sum\limits_{i=1}^m} \pi d_i \delta_{a_i}
\hspace{5em}\mbox{weakly in }{ M}
\]
and for every $\rho > 0$
\[
\int_{\T^2} { E}^\e[v^\e] dx \le
m  (\pi \log{\frac{1}{\rho}} + I(\e, \rho) ) + W(a,d)
+ C \rho + o(1)
\]
as $\e\rightarrow 0$.
\label{construct.data}\end{proposition}
\begin{remark} The condition $[J {v^\e}] \rightharpoonup \sum \pi {d_i} {\delta_{a_i}}$ weakly in $M$ implies the energy lower bound
\begin{equation*}
{\int_{\T^2}} {E^\e}(\ue) dx \geq m \left( \pi \log \frac{1}{\rho} + I(\e,\rho)\right)
+ W(a,d)
+ {\widetilde{C}} \rho + o(1)
\end{equation*}
for some constant ${\widetilde{C}}$ and every $\rho>0$ satisfying \eqref{small.rho}. This follows from Theorem 2 below. 
Therefore, the functions $v_\e$ of Proposition 4 are asymptotically energy minimizing, for the given configuration of vortices.
\label{lower}
\end{remark}

\section{Concentration and structure}

This section identifies two hypotheses on a function $\ue$ which imply
detailed structural properties of $\ue$ as $\e \rightarrow 0$. The first
hypothesis is that $[J \ue ]$ is somewhat close to a weighted sum of Dirac
masses. The second is an energy upper bound consistent with the number of 
Dirac masses. Theorem 1 says that these hypotheses imply the measures
$[J \ue ]$ and ${\mu^\e_{\ue}}$ concentrate to Dirac masses and that $\ue$ is 
uniformly regular away from the points of concentration. A more stringent 
energy upper bound on $\ue$ controls the current $j(\ue)$ and $D|\ue|$ as 
stated in Theorem 2. The proofs, which appear in \cite{ColJer97}, rely
heavily on techniques developed in \cite{Jer95.2}. Similar techniques
have been used by Sandier \cite{San97}.

\begin{theorem}{\bf{(Global structure)}}
Suppose that $u\in H^1({\T}^2; {\R}^2)$,
and that there exist points $x_1, ... , x_m \in {\T}^n$,
integers $d_1, ... , d_m \in  \{ \pm 1 \}$, and
$\e \le r := \frac 14 \min_{i\ne j}|x_i - x_j|$
such that
\begin{equation}
|| \; [Ju] -  \pi \sum_{i=1}^m  d_i \delta_{x_i} \; ||_{{M}^1({\T^2})}
\le \frac{\pi}{200} r,
\label{GS1}\end{equation}
and
\begin{equation}
\int_{{\T}^2} {E}^\e(u) \; dx \le \pi m \log \left(\frac r\e\right) + {\gamma_1}
\label{GS2}\end{equation}
for some constant $\gamma_1$.
Then there exists points $ a_i \in B_{r/2}(x_i)$,
$i=1, ..., m$ such that
\begin{equation}
|| \mu^\e_u - \pi{\sum\limits_{i=1}^m} \delta_{a_i} ||_{ {{M}^1}({\T^2}) } \le 
{O_{\gamma_1}}\left(\frac{1}{|\log \e|}\right),
\label{GS3}\end{equation}
\begin{equation}
|| [Ju] - \pi {\sum\limits_{i=1}^m} d_i \delta_{ a_i} ||_{ {{M}^1}({\T^2}) } \le 
{o_{\gamma_1}}(1),
\label{GS4}\end{equation}
Moreover, for $\rho$ fixed and $0 < \rho< \frac{r}{2}$,
\begin{equation}
\int_{ {\T^2_\rho} }
{{E}^\e}(u) dx \le {O_{\rho, \gamma_1}}(1).
\label{GS5}\end{equation}
\begin{equation}
{{\|Du\|}_{{L^2}({\T^2_\rho})}} \leq {O_{\rho, {\gamma_1} }}(1),
\label{GS5.1}
\end{equation}
Finally for any $p\in [1,2)$, 
\begin{equation}
|| Du ||_{L^p(\T^2)} \le {O_{p, {\gamma_1}}}(1)
\label{GS6}\end{equation}
and
\begin{equation}
|| j(u) ||_{L^p(\T^2)} \le {O_{{p}, {\gamma_1}}}(1)
\label{GS7}\end{equation}
\label{T.GS}\end{theorem}

We have some remarks before commenting on the proof.
\begin{remark} The estimate \eqref{GS1} says $[Ju]$ is $O(1)$-close to a weighted sum of Dirac masses located at the points $x_i$. The energy upper bound \eqref{GS2} allows us to locate points $a_i$ near $x_i$ and conclude $[Ju]$ is $o(1)$-close to a weig
hted sum of Dirac masses located at the points $a_i$. This observation is exploited in the proof of Theorem 3 below.
\label{move}
\end{remark}
\begin{remark} Theorem 1 implies several other useful estimates. For $u$ satisfying the hypotheses of the theorem, we have by \eqref{GS2} that
\begin{equation}
{{\| {{|u|}^2} - 1 \|}_{{L^2}({\T^2})}} \leq C \e {{\left(\log \frac{r}{\e}+1\right)}^\half}.
\label{unear1}
\end{equation}
Interpolating with \eqref{GS6} gives for any $p < \infty$
\begin{equation}
{{\|u\|}_{{L^p}({\T^2})}} \leq {C_p}.
\label{uinLp}
\end{equation}
For any fixed $\rho$ with $0<\rho< \frac{r}{2}$, the estimate \eqref{GS5.1} implies
\begin{equation}
{{\|Ju\|}_{{L^1}({\T^2_\rho})}} \leq {C_\rho}.
\label{JuinL1}
\end{equation}
\label{estimates}
\end{remark}
\begin{remark}
Related results for vector fields from ${\R^n} \longmapsto {\R^n}$ generalizing the two dimensional results above appear in \cite{ColJer97}. 
\label{vector}
\end{remark}

We briefly describe the idea of the proof of Theorem 1 appearing in \cite{ColJer97}. The theorem follows from a localized version with the hypotheses
\begin{equation*}
{{\left\| [Ju] - \pi d {\delta_{x_i}} \right\| }_{{M^1}({B_r({x_i})})}} \leq {\frac{\pi}{200}} r
\end{equation*}
\begin{equation*}
{\int_{{B_r}({x_i})}} {E^\e}(u) dx \leq \pi \log \frac{r}{e} + {\gamma_1}.
\end{equation*}
The Jacobian condition implies $u$ must exhibit
nonzero degree in many subsets of $B_r$.
One way this condition can be satisfied, which is consistent with
the assumed upper bound, is if $u$ has a single isolated
vortex of degree $d$ near the center $x_i$. Careful lower bounds
show that, if $u$ deviates significantly from this description ---
for example, if $u$ has more than one vortex, or if the energy
or vorticity of $u$ is spread out too much --- then the total
energy of $u$ would be too large, contradicting the energy
bound. The main technical point is thus lower bounds.
These are proved by identifying some set which contains
all concentrations of vorticity, and then covering it by balls
in some optimal fashion. One can use this covering to relate the
energy of the function to the distribution of vorticity.
As noted above, this sort of argument has appeared before in
\cite{Jer95.2} and in the work of Sandier \cite{San97}.

The next result provides additional information about $\ue$ under a stronger energy upper bound hypothesis.

\begin{theorem}
Suppose that $\ue \in H^1$ is a sequence such that
\begin{equation}
[J u^\e] \rightharpoonup \pi \sum_{i=1}^m d_i \delta_{a_i}~~{\text{weakly as measures}},
\label{hyp1.1a}\end{equation}
and that there exists some $\gamma_2 \geq 0$ such that
\begin{equation}
\int {E}^\e (\ue) dx \le  m  \left( \pi\log{\frac{1}{\rho}} + 
I(\e, \rho) \right) + W(a,d) + C\rho + \gamma_2 + o(1)
\label{oone.data1a}\end{equation}
as $\e \rightarrow 0$, for every $\rho > 0$.
Then 
\begin{equation}
\limsup_{\e \rightarrow 0} || \frac 1{|\ue|} j(\ue)
- j(H) ||^2_{L^2(\T^2_\rho)} \le C\gamma_2
\label{conclusion1}\end{equation}
and
\begin{equation}
\limsup_{\e\rightarrow 0} || \; D|\ue| \; ||^2_{L^2(\T^2_\rho)} \le
C\gamma_2
\label{conclusion2}\end{equation}
for every $\rho > 0$.
Here $H = H(\cdot; a,d)$ is the canonical harmonic map and $C$ is a
universal constant.
\label{T.renorm}\end{theorem}

\begin{remark}
Observe that the conclusion of the theorem is ridiculous if $\gamma_2$ is negative which implies the claim in Remark 4 .
\label{ridiculous}
\end{remark}

We conclude this section with a brief discussion of the $m=0$
case. Consider $GLS_\e$ with initial data $\phie = {\rho^\e} {e^{i
{\theta}}}$ with ${\rho^\e}, {\theta} \in \R$. The solution is
\begin{equation*}
\ue(x,t) = {\rho^\e} {e^{i\left[{\theta} + {\frac{1}{\e^2}} \left(
{{({\rho^\e})}^2} -1 \right) t \right]}}.
\end{equation*}

If $|
{{({\rho^\e})}^2} -1 | > O({\e^\alpha})$ for $\alpha < 2$ then
$\ue$ oscillates rapidly in the $t$ variable and $\ue \rightharpoonup
0$ weakly in ${L^2}(dxdt)$ as $\e \rightarrow 0$. The canonical
harmonic map in this case is $H(x) = {e^{i {\theta_1}}}$ for some
constant ${\theta_1} \in [0, 2 \pi)$. The rapid temporal oscillations
require $|
{{({\rho^\e})}^2} -1 | = O({\e^2})$ to have $\ue \rightarrow H$
in any strong sense.

The assumptions \eqref{GS2} and $m=0$ imply $|
{{({\rho^\e})}^2} -1 | = O({\e})$. Contrast this with assumption
\eqref{oone.data1a} under the conditions ${\gamma_2}=0$ and $m=0$ which yields
 $|{{({\rho^\e})}^2} -1 | = o({\e})$.

The quantity $j(u)$ is insensitive to the temporal oscillations in $u$
since $j(u) = j({e^{i \alpha}}u)$ for $\alpha \in \R$. It is therefore
natural to expect $j(u)$ to retain more information in the $\e
\rightarrow 0$ limit. This expectation is verified later in Theorem 3(iv).

\section{Vortex paths and convergence}

Theorem 1 describes a function ${\ue}$ (through the measures $[J\ue]$
and ${\mu_u^\e}$~) provided that $\ue$ satisfies the hypotheses \eqref{GS1}
and \eqref{GS2}. The theorem identifies Dirac masses as the $\e \rightarrow
0$ limits of $[J\ue],~ {\mu_u^\e}$. The hypotheses \eqref{GS1} and \eqref{GS2}
 are chosen to allow for a similar description of $\ue(t)$, the solution of
${GLS}_\e$ with initial data $\ue(0) = \phie$, at each $t$ in a time
interval $[0,T)$. In this section, we exploit the identity \eqref{wjacid}
and the results stated in Section 3 to show the Dirac masses in Theorem 1
move along Lipschitz paths provided $\ue$ solves ${GLS}_\e$. The paths
are characterized as solutions of certain ordinary differential
equations in the next section under stronger assumptions on the
initial data.

We assume the initial data $\phie \in {H^1}({\T^2}; {\R^2})$ satisfies the hypotheses \eqref{JtoDirac} and \eqref{Ieupper}. So, there exist distinct points $\{ {\alpha_1}, \dots, {\alpha_m} \} \subset {\T^2}$, integers $\{{d_1}, \dots,{d_m} \}$ with ${d_i
}=\pm 1$ and $\sum {d_i} =0$, fixed independently of $\e$ such that

\begin{equation}
[J \phie ] \rightharpoonup \pi {\sum_{i=1}^m} {d_i} {\delta_{{\alpha_i}}} 
\label{JDirac}
\end{equation}
\begin{equation}
{\int\limits_{{\T^2}}} {E^\e}(\phie ) ~dx \leq \pi m \log \frac{1}{\e} + {\gamma_1},
\label{E1upper}
\end{equation}

We will also assume
\begin{equation}
{\int_{\T^2}} j(\phie) dx = 0.
\label{phasezero}
\end{equation}
In light of \eqref{jev}, this implies $\int j(\ue(t))dx = 0$ for all time. 
This assumption is not essential but simplifies the exposition. There are 
modifications to the statements of Theorems 3 and 4 if 
$\int j(\ue(t))dx \neq 0$ which appear in \cite{ColJer97}.

For $\e>0$, Bourgain's recent work \cite{B1} shows ${GLS}_\e$ is globally (in time) wellposed below $H^1$. More precisely, there exists ${s_0} < 1$ such that for all $s>{s_0}$ there is a uniquely defined continuous map
\begin{equation*}
{H^s} \longmapsto X \subset C([0,\infty); {H^s})
\end{equation*}
taking
\begin{equation*}
\phie \longmapsto \ue ,
\end{equation*}
where $\ue$ is the solution of ${GLS}_\e$. 

We can now state our first main theorem.

\begin{theorem}
Let $\ue$ be the solution of ${GLS}_\e$ with initial data $\phie$ satisfying \eqref{JDirac} and \eqref{E1upper} and \eqref{phasezero}. Then, after passing to a subsequence as $\e \rightarrow 0$, there exists a $T>0$ (independent of $\e$) and Lipschitz pat
hs ${a_i}:[0,T)\longmapsto {\T^2},~{a_i}(0) = {\alpha_i}$ such that \newline
(i) $[J \ue (t) ] \rightharpoonup \pi {\sum\limits_{i=1}^m} {d_i} {\delta_{{a_i}(t)}}$ weakly as measures for all $t \in [0,T)$. \newline
(ii) ${\mu_{\ue (t)}^\e} \rightharpoonup \pi {\sum\limits_{i=1}^m} {\delta_{{a_i}(t)}}$ weakly as measures for all $t \in [0,T)$.\newline
(iii) ${{|\ue(t)|}^2} \rightarrow 1 $ in ${L^2}(dx)$ for all $t \in [0,T)$.\newline
(iv) $j(\ue) = (i \ue)\cdot D \ue \rightharpoonup j(H))$ weakly in ${L^p}(dx dt)$
for all $1 \leq p < 2$ where $H(\cdot ,t)=H(\cdot,a(t),d)$ is the canonical harmonic map with singularities of degree $(d)$ located at $(a(t))$.
\end{theorem}
\begin{proof}
1. Define for $r= {\min_{i \neq j}} \frac14 |{\alpha_i} - {\alpha_j}|$ with $0 < \e \ll r \leq 1$ the quantity
\begin{equation*}
{T^\e} = \sup \left\{ { t\geq 0:~{{\left\| [J \ue (s) ] - [J \phie ]  \right \|}_{{M^1}({\T^2})}} \leq \frac{\pi}{400} r~{\text{for all}}~0\leq s \leq t } \right\} .
\end{equation*}
Recall that $Ju=\det Du =
{u^1_{{x_1}}}{u^2_{{x_2}}}-{u^1_{{x_2}}}{u^2_{{x_1}}}$. The estimates
\begin{equation*}
{{\left\| [J\ue] - [J\phie ]\right\|}_{M^1}} \leq {{\left\| J\ue -
J\phie \right\|}_{{L^1}}} \leq C {{\left\| \ue - \phie \right\|
}_{{H^1}}} {{\| \ue + \phie \|}_{{H^1}}}
\end{equation*}
and continuity of the flow of $\phie \longmapsto \ue(t)$ through $H^1$ guarantee ${T^\e} > 0$. \newline
2. {\bf{Claim:}} For $s,t$ satisfying $0 \leq s,t \leq {T^\e}$ we have
\begin{equation*}
{{\left\| [J \ue (s) ] - [J \ue (t) ]  \right \|}_{{M^1}({\T^2})}} \leq c|s-t| + {o_{\gamma_1}}(1).
\end{equation*}
{\it{Proof of Claim:}} The definition of $T^\e$ guarantees for all $t \in [0,{T^\e})$ that the hypothesis \eqref{GS1} of Theorem 1 holds if $\e$ is sufficiently small. Therefore, for each $t \in [0, {T^\e})$, we can find points ${{a_i}(t)} \in {B_{{\frac{
r}{2}}}}({\alpha_i}),~  i=1,\dots,m$, for which
\begin{equation}
{{\left\| [J \ue (t) ] - \pi {\sum\limits_{i=1}^m} {d_i} {\delta_{{a_i}(t)}}  \right \|}_{{M^1}({\T^2})}} \leq {o_{\gamma_1}}(1),
\label{JooneDirac}
\end{equation}
by \eqref{GS4}. Of course, the ${{a_i}(t)}$ may depend upon $\e$.
So we can estimate
\begin{equation*}
{{\left\| [J \ue (s) ] - [J \ue (t) ]  \right \|}_{{M^1}({\T^2})}} \leq 
{{\left\| \pi {\sum\limits_{i=1}^d} {d_i}  ({\delta_{{a_i}(s)}}  -   {\delta_{{a_i}(t)}})   \right \|}_{{M^1}({\T^2})}} + {o_{\gamma_1}}(1).
\end{equation*}
By \eqref{bcl}, we can estimate by
\begin{equation*} 
\leq \pi {\sum\limits_{i=1}^m} |{a_i}(s)- {a_i}(t)| + {o_{\gamma_1}}(1).
\end{equation*}
The claim will be established once we show for $i=1, \dots, m$,
\begin{equation}
|{a_i}(s)- {a_i}(t)| \leq c|s-t| + {o_{\gamma_1}}(1).
\label{vortlip}
\end{equation} 
3. We prove \eqref{vortlip} by using the identity \eqref{wjacid}. Fix $i$ and observe that ${{a_i}(s)}, ~{{a_i}}(t) \in {B_{{\frac{r}{2}}}}({\alpha_i})$ for all $s,t \in [0,{T^\e})$. There exists an $\eta \in {C^\infty_c}({B_r}({\alpha_i}))$ satisfying
\begin{equation*}
\eta(x) = \nu \cdot x ~~{\text{for}}~ x \in {B_{{\frac{3r}{4}}}}({{a_i}}(0)),~\nu \in {S^1}
\end{equation*}
and
\begin{equation*}
\pi|{a_i}(s)- {a_i}(t)|=\pi {d_i} \int \eta ({\delta_{{a_i}(s)}}  -   {\delta_{{a_i}(t)}}).
\end{equation*}
The conditions on $\eta$ guarantee that $supp({D^2}\eta) \subset {B_r}({\alpha_i}) \setminus {B_{{\frac{3r}{4}}}}({\alpha_i})$. Notice that $\eta$ depends upon the index $i$.

Insert the function $\eta$ described above into \eqref{wjacid} and again use \eqref{JooneDirac} to observe
\begin{equation*}
\pi |{a_i}(s)- {a_i}(t)|={\int\limits_s^t}
{\int\limits_{{B_r}({\alpha_i})}} {\eta_{{x_j}{x_l}}} {\J_{jk}}~
{u_{{x_k}}} \cdot {u_{{x_l}}}~dx~d\tau + {o_{\gamma_1}}(1). 
\end{equation*}
The support properties of ${\eta_{{x_j}{x_l}}}$ permit us to replace $
{B_r}({\alpha_i})$ by ${B_r}({\alpha_i}) \setminus {B_{{\frac{3r}{4}}}}({\alpha_i})$. Finally, we estimate by
\begin{equation*}
\leq  |s-t|~ {{\| {D^2} \eta \|}_{L^\infty}} {\sup_{\tau \in [s,t]}} {{\|
Du(\tau) \|}_{{L^2}({B_r} \setminus {B_{\frac{3r}{4}}})}}.
\end{equation*}
The size of ${{\| {D^2} \eta \|}_{L^\infty}}$ depends upon $r$ but is
independent of $\e$ and \eqref{GS5.1} permits us to control the $Du$ term by a
constant independent of $\e$, so \eqref{vortlip} follows and the claim is
proven. We also note that the
claim implies ${T^\e}$ may be taken independently
of $\e$, so we denote this quantity by $T$ from now on. \newline
4. The remaining convergence claims follow from the bounds stated in
Theorem 1 and passing to subsequences, except for (iii) which follows 
directly from (4.2). We prove (iv). Fix any $p \in
[1,2)$. Since the conditions of Theorem 1 hold for every $t \in [0,T)$,
we deduce from \eqref{GS7} that 
\begin{equation*}
{{\| j(\ue) \|}_{{L^p}({\T^2} \times [0,T))}} \leq {O_{p,{\gamma_1},T}}(1).
\end{equation*}
It follows, upon passing to a subsequence as $\e \rightarrow 0$, that
\begin{equation*}
j(\ue) \rightharpoonup {\bar{j}} ~~~~{\text{weakly in }}~{L^p}(dx dt)
\end{equation*}
for some ${\bar{j}}$. We wish to identify ${\bar{j}}$.

Let $\phi \in {C^\infty_0}({\T^2} \times [0,T))$. The identity \eqref{L2ev}
implies
\begin{eqnarray*}
\int j(\ue) \cdot D \phi dxdt &=& \int {\phi_t} {\frac{{|\ue|}^2}{2}} dx dt\\ 
&=& \int {\phi_t} \half {{({{|\ue|}^2} -1)}} dxdt \\
& \rightarrow& 0 \\
\end{eqnarray*}
as $\e \rightarrow 0$ for every $t$, by (iii). Therefore $\div {\bar{j}} = 0$.
Moreover, from (i) we have
$\nabla \times {\bar{j}} = 2[{\bar{J}}] \otimes dt = 2 \pi \sum {d_i}
{\delta_{{a_i}(t)}} \otimes dt$ weakly.

Let $H(x,t) = H(x,a(t),d)$. If we define $V= {\bar{j}} - j(H)$, we have
\begin{equation*} 
\div V = \nabla \times V = 0
\end{equation*}
weakly.
Let $\eta^\delta$ be a standard mollifier and set ${V^\delta} = V *
{\eta^\delta}$. The convolution here is in space and time. The above
considerations imply
\begin{equation*}
\div {V^\delta} = \nabla \times {V^\delta} = 0
\end{equation*}
in ${\T^2}$ for every $t < T$. Since ${V^\delta}$ is smooth, this
implies ${V^\delta}(x,t) = {g^\delta}(t)$. Letting $\delta \rightarrow
0$, we find that $V$ is also constant in $x$ for each fixed $t$. For
any fixed $t$ we have
\begin{eqnarray*}
\int V(x,t) dx &=& \int \left( {\bar{j}} - j(H) \right) (x,t) dx \\
&=& \int {\bar{j}} (x,t) ~dx \\
&=& {\lim\limits_{\e \rightarrow 0}} \int j(\ue) (x,t) dx  \\
&=& 0 
\end{eqnarray*}
using \eqref{phasezero}.
\end{proof}

\begin{remark}
The proof given above may be iterated until
\begin{equation*}
T=\inf\{t>0: |{a_i}(t)-{a_j}(t)| \rightarrow 0 ~~{\text{for some}}~~ i \neq j \}.
\end{equation*}
\label{iterated}
\end{remark}

\section{Vortex dynamics}

Under the more restrictive upper bound on the energy \eqref{oone.data1a}, we characterize the vortex paths ${a_i}(\cdot)$ as solutions of a system of ordinary differential equations that conserves the renormalized energy. More precisely, we have the follo
wing result.

\begin{theorem}
Suppose ${\phie}$ satisfies
\begin{equation}
[J\phie] \rightharpoonup \pi {\sum\limits_{i=1}^m} {d_i} {\delta_{{\alpha_i}}} 
~~{\text{weakly as measures}},
\end{equation}
and for every $\rho > 0$ as $\e \rightarrow 0$,
\begin{equation}
{\int_{\T^2}} {E^\e}({\phi^\e})dx \leq m \left( \pi  \log
{\frac{1}{ \rho}} + I(\e,\rho) \right) + W(\alpha,d) + C \rho + o(1). 
\label{Etite}
\end{equation} 
Let $\{{a_1}(\cdot), \dots, {a_m}(\cdot) \}$ and $H$ be as in Theorem 3. Then the following statements hold:\newline
(i) For each $i$ and for $t \in [0,{T_1})$, 
\begin{equation*}
\left\{ \matrix
{\frac{d}{dt}} {a_i} = 2 {\sum\limits_{j:j \neq i}} {d_j} \nabla \times F({a_i}-{a_j}) = -{\frac{1}{\pi}} {d_i} \J {D_{a_i}} W({a}, d) \\
{a_i}(0) = {\alpha_i}.
\endmatrix
\right.
\end{equation*}
Here ${T_1}>0$ can be taken to be the largest time such that the above ODE has a solution on $[0,{T_1})$.
\newline
(ii) For every $t\in [0,{T_1})$,
\begin{equation*}
{\frac{1}{|\ue |}} j(\ue) \rightarrow j(H) ~~~{\text{strongly in }} {L^2_\loc}({\T^2} \setminus (a(t)) ).
\end{equation*}
(iii) For every $\rho > 0$ and $t \in [0,{T_1})$,
\begin{equation*}
{\lim\limits_{\e \rightarrow 0}}~~~ {\min\limits_{\alpha \in [0,2 \pi)}} 
{{\left\|
\ue(\cdot,t) - {e^{i\alpha}} H(\cdot,a(t),d) \right\| }_{{H^1}({\T^2_\rho})}} = 0
\end{equation*}

\end{theorem}

\begin{remark} For an example of a point vortex system 
which develops a singularity in finite time, see 
Marchioro and Pulvirenti \cite{MarPul94}. We expect, however, that
for generic initial data the ODE does not develop singularities,
in which case the above result is valid globally in time.
\end{remark}

\begin{remark} The example given at the end of Section 3 shows 
that the sort of convergence in $(iii)$ is more or less optimal,
unless \eqref{Etite} is strengthened still further.
\end{remark}

\begin{proof}1. Let ${a_i}(t),~ i =1, \dots, m$ denote the paths of Theorem 2.
Let ${b_i}(t)$ denote the solution of the system
\begin{equation}
\left\{ \matrix
{\frac{d}{dt}} {b_i} = 2 {\sum\limits_{j:j\neq i}} {d_j} \nabla \times F({b_i}-{b_j}) \\
{b_i}(0) = {\alpha_i}.
\endmatrix
\right.
\label{bdyna}
\end{equation}
A calculation shows that
\begin{equation*}
{D_{a_i}} W = -2 \pi {\sum\limits_{j:j \neq i}} {d_i}{d_j} DF({a_i}-{a_j}).
\end{equation*}
Therefore, the ODE in \eqref{bdyna} may be reexpressed, as in (i), in Hamiltonian form showing that the renormalized energy $W$ is conserved.
This ODE system has a unique solution on a nontrivial time interval $[0,{T'})$. Let
\begin{equation*}
{T_1} = \min(T, {T'})
\end{equation*}
where $T$ is as in Remark \ref{iterated}. 
Note that $T_1$ is independent of $\e$. 
We wish to show for all $i$ that ${b_i}(t)$ coincides with ${a_i}(t)$ on the time interval $[0, {T_1})$. Observe that this will imply ${T_1}={T'}$.

For $t \in [0, {T_1})$,
let 
\begin{equation*}
\zeta(t) = {\sum\limits_i} |{b_i}(t) - {a_i}(t)|.
\end{equation*}
It suffices to prove that, given any ${\widetilde{T}} < {T_1}$, we can find
some small $\delta(\Ttilde) > 0$ and a constant $C= C(\Ttilde)$ such that
\begin{equation}
{\frac{d}{dt}}\zeta(t) \leq C \zeta(t)
\label{gron}
\end{equation}
for $a.e.~t \in [0,{\Ttilde}]$ whenever $\zeta(t) \leq \delta$. We will show
that \eqref{gron} holds at each point where ${a_i}(\cdot)$ is differentiable for all $i$; by Rademacher's theorem, this condition is satisfied on a set of full measure.

Fix $\Ttilde < {T_1}$. By Remark \ref{iterated}, 
there is some $r=r(\Ttilde)> 0$ for which
\begin{equation}
{{\min\limits_{i\neq j,~t\leq \Ttilde}}} |{a_i}(t) - {a_j}(t)| \geq 4r.
\label{asep}
\end{equation}
\newline
2. We use the fact that ${b_i}$ solves \eqref{bdyna} and the triangle inequality to estimate
\begin{eqnarray*}
{\frac{d\zeta}{dt}} &\leq &{\sum\limits_i} |{b_{i,t}} - {a_{i,t}}| \\
&\leq & 2 {\sum\limits_i} \left| {\sum\limits_{j: j\neq i}} {d_j} \nabla \times F({b_i}-{b_j}) - {\sum\limits_{j: j\neq i}} {d_j}\nabla \times F({a_i}-{a_j}) \right| \\
& + &{\sum\limits_i} \left| {a_{i,t}} - 2 {\sum\limits_{j: j\neq i}} {d_j}\nabla \times F({a_i}-{a_j}) \right| \\
&= &Term~ 1 + Term~ 2.
\end{eqnarray*}

We immediately dispose of $Term ~1$. Fix $s< \Ttilde$ and a pair of indices $ i \neq j$. Let $h=|({b_i}(s)-{b_j}(s)) - ({a_i}(s) - {a_j}(s))|.$ Note that by assumption $h \leq \zeta(s) \leq \delta$. By Taylor's theorem, at the fixed time $s$, we have
\begin{eqnarray*}
|\nabla \times F({b_i}-{b_j}) - \nabla \times F({a_i}-{a_j})| &\leq &h {\max\limits_{\{x: |x-({a_i}-{a_j})| \leq h \}}} |{D^2}F| \\
&\leq & C \zeta(s).
\end{eqnarray*}
The last inequality follows from \eqref{asep} provided $\delta < r$. Therefore, $Term~1$ satisfies the desired estimate \eqref{gron}.
\newline
3. We turn our attention to $Term~ 2$, which is a sum of terms
$(Term~2)_i$, with $i=1, \dots, m$. Suppose that each function
${a_i}(\cdot)$ is differentiable at $s \in [0, \Ttilde)$. Fix $\eta
\in {C^\infty_0}$ such that 
$supp(\eta) \subset {B_{r(\Ttilde)}}({a_i}(s))$ and $\eta(x) = \nu \cdot x$ in a neighborhood of ${a_i}(s)$. Here we take $\nu \in {S^1}$ to satisfy
\begin{equation}
{{(Term~2)}_i} = {d_i} \nu\cdot \left( {a_{i,t}}(s) - 2{\sum\limits_{j:j\neq i}} {d_j} \nabla \times F( {a_i}(s) - {a_j}(s) ) \right).
\label{TermTwo}
\end{equation}
We will write $\en$ for an appropriate subsequence as $\e \rightarrow 0$.
Since
\begin{equation*}
[J \uen (t) ] \rightharpoonup \pi \sum {d_i}
{\delta_{{a_i}(t)}}~~~{\text{weakly as measures}} 
\end{equation*}
we can rewrite
\begin{eqnarray*}
{d_i} \nu \cdot {a_{i,t}}(s) & = & {\lim\limits_{h \rightarrow 0}} {d_i}{\nu}
\cdot \frac{1}{h} ({a_i}(s+h) - {a_i}(s)) \\
&=& {\lim\limits_{h \rightarrow 0}}~  {\lim\limits_{n \rightarrow \infty}}  
{\frac{1}{\pi h}} {\int_{\T^2}} \left( \eta [J \uen (s+h) ] - \eta [J \uen (s) ] \right) dx \\
& = & {\lim\limits_{h \rightarrow 0}} ~ {\lim\limits_{n \rightarrow \infty}}  
{\frac{1}{\pi h}} {\int\limits_s^{s+h}} {\int_{\T^2}} {\eta_{{x_j}{x_l}}} 
{\J_{jk}} {u^{\en}_{x_k}}\cdot{u^{\en}_{x_l}} dx dt.
\end{eqnarray*}
We used \eqref{wjacid} in the last step.

Let $H(x,t) = H(x, a(t), d)$. We reexpress the remaining term in \eqref{TermTwo}
using Lemma 1 which is stated and proven below,
\begin{eqnarray*}
&&{d_i} \nu \cdot \left( 2 {\sum\limits_{j:j\neq i}} {d_j} \nabla \times F ( {a_i}(s) - {a_j}(s)) \right) \\
&= &{\lim\limits_{h \rightarrow 0}} \frac{2}{h}{\int\limits_s^{s+h}}
{d_i} \nu \cdot 
\left( {\sum\limits_{j:j\neq i}}{d_j} \nabla \times F ( {a_i}(t) - {a_j}(t)) \right) dt \\
&=&{\lim\limits_{h \rightarrow 0}} \frac{1}{\pi h}{\int\limits_s^{s+h}} {\int_{\T^2}} {\eta_{{x_j}{x_l}}}{\J_{jk}} {j^k}(H)  {j^l}(H) dx dt.
\end{eqnarray*}
Therefore,
\begin{equation}
{{(Term~2)}_i} = {\lim\limits_{h \rightarrow 0}}~
{\lim\limits_{n \rightarrow \infty}} \frac{1}{\pi h}{\int\limits_s^{s+h}} {\int_{\T^2}} {\eta_{{x_j}{x_l}}}{\J_{jk}} \left({u^{\en}_{x_k}}\cdot{u^{\en}_{x_l}} - {j^k}(H){j^l}(H) \right) dx dt.
\end{equation}
Inside the integral, $H= H(\cdot, a(t), d)$ and $D\ue = D\ue (t)$. 

On any set where $|u| > 0$, we can decompose ${u_{x_k}}$
\begin{equation*}
{u_{x_k}} = \left[ \frac{iu}{|u|} \cdot {u_{x_k}} \right] \frac{iu}{|u|}
+ \left[ \frac{u}{|u|} \cdot {u_{x_k}} \right] \frac{u}{|u|}
\end{equation*}
\begin{equation}
= \frac{{j^k}(u)}{|u|}  \frac{iu}{|u|} + {{|u|}_{x_k}} \frac{u}{|u|}.
\label{Dudec}
\end{equation}
Since $|Du|=0~a.e.$ on the set $|u|=0$, the representation above holds
$a.e.$ on $supp(Du)$ and
\begin{equation}
{u^{\e}_{x_k}}\cdot{u^\e_{x_l}} = {\frac{1}{{|\ue|}^2}} {j^k}(\ue) {j^l}(\ue) + 
{{|\ue|}_{x_k}} {{|\ue|}_{x_l}}. 
\end{equation}
We will have proved \eqref{gron} when we show
\begin{equation}
{\lim\limits_{h \rightarrow 0}}~ {\lim\limits_{n \rightarrow \infty}}  \frac{1}{h}{\int\limits_s^{s+h}} {\int_{\T^2}} {\eta_{{x_j}{x_l}}}{\J_{jk}} 
\left({{|\uen|}_{x_k}} {{|\uen|}_{x_l}} \right) dx dt \leq C \zeta(s),
\label{fiveeight}
\end{equation}
and
\begin{equation}
{\lim\limits_{h \rightarrow 0}}~{\lim\limits_{n \rightarrow \infty}}   \frac{1}{h}{\int\limits_s^{s+h}} {\int_{\T^2}} {\eta_{{x_j}{x_l}}}{\J_{jk}} \left( {\frac{1}{{|\uen|}^2}} {j^k}(\uen) {j^l}(\uen)
- {j^k}(H) {j^l}(H) \right) dx dt \leq C \zeta(s).
\label{fivenine}
\end{equation}
These estimates will follow from the tight upper bound \eqref{Etite} (see also \eqref{oone.data1a})
on the energy and energy conservation.
\newline
4. The renormalized energy $W$ is conserved for solutions $b(\cdot)$ of \eqref{bdyna}. This is apparent from the fact that \eqref{bdyna} can be written in Hamiltonian form. Also ${\int_{\T^2}} {E^\e}(\ue(\cdot,t))(x) dx$ is conserved for solutions $\ue$ o
f $GLS_\e$. Therefore, for every $t \leq \Ttilde$ and every $\rho > 0$, the upper bound \eqref{Etite} gives
\begin{eqnarray*}
{\int_{\T^2}} {E^\e}(\ue(\cdot,t))(x) dx &=& {\int_{\T^2}} {E^\e}(\phie(\cdot))(x) dx \\
&\leq & m( \pi \log {\frac{1}{\rho}} + I(\e, \rho)) + W(a(0),d) + C \rho +o(1) \\
&= & m (\pi \log {\frac{1}{\rho}} + I(\e, \rho)) + W(b(t),d) + C \rho +o(1).
\end{eqnarray*}
Arguing as in the estimate of $Term~1$, we see that 
\begin{equation*}
W(b(t)) - W(a(t)) \leq C \sum |{b_i}(t) - {a_i}(t)| = C \zeta(t)
\end{equation*}
provided $\delta $ is small enough. Therefore
\begin{equation}
{\int_{\T^2}} {E^\e}(\ue(\cdot ,t))(x) dx \leq m \left(\pi \log \frac{1}{\rho} + I(\e,\rho)\right) + W(a(t), d) + C\rho + C \zeta(t) + o(1),
\label{Ezeta}
\end{equation}
as $\e \rightarrow 0$ for every $\rho > 0$. We have from Theorem 3 that
\begin{equation}
[J \uen (t) ] \rightharpoonup \pi {\sum\limits_{i=1}^m} {d_i} {\delta_{{a_i}(t)}} ~~~{\text{weakly as measures}}.
\label{Jgoes}
\end{equation}

The conditions \eqref{Ezeta}, \eqref{Jgoes} are precisely the hypotheses of Theorem 3
with ${\gamma_2} = C \zeta(t)$. So, for every $t \in [s,s+h]$,
\begin{equation}
{\limsup\limits_{n\rightarrow \infty}} {{\left\| D {{|\uen|}}(\cdot,t) \right\|}_{{L^2}({\T^2_\rho})}^2} \leq C \zeta(t)
\label{Modbound}
\end{equation}
and
\begin{equation}
{\limsup\limits_{n\rightarrow \infty}} {{\left\| {\frac{1}{|\uen|}} j(\uen) - j(H) 
 \right\|}_{{L^2}({\T^2_\rho})}^2} \leq C \zeta(t)
\label{jconv}
\end{equation}
These estimates allow us to prove \eqref{fiveeight}, \eqref{fivenine}.

We quickly estimate \eqref{fiveeight} by observing
\begin{equation}
\eqref{fiveeight} \leq {\lim\limits_{h \rightarrow 0}}~{\lim\limits_{n \rightarrow \infty}} {{\| {D^2} \eta \|}_{{L^\infty}({\T^2_\rho})}} \frac{C}{h} {\int\limits_s^{s+h}} {{\left\| D {{|\uen|}}(\cdot,t) \right\|}_{{L^2}({\T^2_\rho})}^2} dt
\end{equation}
and using \eqref{Modbound}. \newline
5. We now establish \eqref{fivenine}. First, we show that
\begin{equation}
\left(\frac{1}{|\uen|} j(\uen) - j(H)\right) \rightharpoonup 0 
~~~{\text{weakly in }} 
{{L^2}({\T^2_\rho} \times [s,s+h])}.
\label{fivetwenty}
\end{equation}
To see this, note that $\frac{1}{|\uen|} j(\uen)$ is uniformly bounded in $
{{L^2}({\T^2_\rho} \times [s,s+h])}$ and hence converges weakly to some limit ${\bar{j}}$. We know from Theorem 3(iv) that $j(\uen) \rightharpoonup j(H)$ weakly in ${L^p}(dxdt)$ for all $1 \leq p < 2$. We also know from Theorem 3(iii) that ${{|\uen|}^2} \
rightarrow 1$ strongly in ${L^2}(dxdt)$. Thus
\begin{eqnarray*}
j(H) &=&~ {\text{weak}}~ {L^1} {\lim\limits_{n \rightarrow \infty}} j(\uen) \\
&=& ~ {\text{weak}}~ {L^1} {\lim\limits_{n \rightarrow \infty}} \left(\frac{j(\uen) }{|\uen|} |\uen| \right) \\
&=& \left( ~{\text{strong}}~ {L^2} {\lim\limits_{n \rightarrow \infty}} |\uen| \right) \left( ~ {\text{weak}}~{L^2} {\lim\limits_{n \rightarrow \infty}}\frac{j(\uen) }{|\uen|}  \right) \\
&=& {\bar{j}}
\end{eqnarray*}
which proves \eqref{fivetwenty}.

For fixed $k,l$, observe that the quadratic term in \eqref{fivenine} can be written
\begin{equation*}
\left( \frac{1}{|\uen|} {j^k}(\uen) - {j^k}(H) \right) \left(\frac{1}{|\uen|} {j^l}(\uen) + {j^l}(H) \right) 
\end{equation*}
and reexpressed as
\begin{equation*}
\left( \frac{1}{|\uen|} {j^k}(\uen) - {j^k}(H) \right) \left(\frac{1}{|\uen|} {j^l}(\uen) - {j^l}(H) \right) + 2 {j^l}(H) \left( \frac{1}{|\uen|} {j^k}(\uen) - {j^k}(H) \right).
\end{equation*}
Since $\left( \frac{1}{|\uen|} {j^k}(\uen) - {j^k}(H) \right) \rightharpoonup 0$ weakly in ${L^2}({\T^2_\rho} \times [s,s+h])$ and ${j^l}(H)$ does not depend upon $n$, the second expression contributes nothing as $n \rightarrow \infty$. The first expressi
on is controlled using \eqref{jconv}. \newline
6. Since we have appropriately bounded $Term ~1$ and $Term~2$, we have
proven \eqref{gron}. Gronwall's inequality implies $\zeta = 0$ which gives
(i) of the Theorem. Since $\zeta = 0$, \eqref{jconv} implies (ii).
We conclude by proving (iii). Fix $t$ and $\rho>0$. Let $\ue$
be a subsequence which converges in ${L^2}({\T^2_\rho})$ to some
limit ${\bar{u}}$. We may assume, by \eqref{GS5.1}, that
$D{\ue} \rightharpoonup D{\bar{u}}$ weakly in ${L^2}({\T^2_\rho})$.
Also, (ii) gives
\[
\liminf_{\e\rightarrow 0} \| D\ue \|_{{L^2}({\T^2_\rho})} \ge
{\lim_{\e \rightarrow 0}}
{{\left\| \frac{j(\ue)}{|\ue|} \right\|}_{{L^2}
({\T^2_\rho})}}
= {{\| D\bar u \|}_{{L^2}({\T^2_\rho})}}.
\]
Therefore $D\ue \rightarrow D{\bar{u}}$ strongly in
${L^2}({\T^2_\rho})$. Finally, since $j({\bar{u}}) = j(H)$,
Proposition 2 implies ${\bar{u}} = {e^{i\alpha}} H$ for some 
$\alpha \in \R$.

\end{proof} 

We state and prove the lemmas employed above.
\begin{lemma} 
Suppose that $\eta \in C^2$ and that
\[
supp(\eta) \cap \{ a_1, ... , a_m \} = \{ a_i \};
\hspace{5em}
D^2\eta \equiv 0\hspace{.5em}
\mbox{in a neighborhood of }a_i.
\]
Let $H := H(\cdot; a,d)$ be the canonical harmonic map.
Then
\[
\int_{\T^2} \eta_{x_j x_l} 
\J_{jk} j^k(H) j^l(H) 		=
d_i D\eta(a_i) \cdot \left(2 \pi \sum_{j:j\ne i} d_j \; \nabla \times F(a_i - a_j)
\right)	
\]
\label{velocity.comp}\end{lemma}

\begin{remark} This computation remains valid if
$d_1, ..., d_m$ assume arbitrary integer values, that is,
if we lift the assumption that $d_i = \pm 1$ for all $i$.
\end{remark}

\begin{proof}
1. We reexpress the integral in the lemma. Recall that $j(H) = - \nabla \times
\Phi$ where $\Phi$ satisfies
\begin{equation}
\Delta \Phi = {\sum_{i=1}^m} 2\pi {d_i} {\delta_{a_i}},
\label{B}
\end{equation}
and, using \eqref{Phi}, we write
\begin{equation}
\Phi(x) = {d_i} F(x-{a_i}) + G(x)
\label{C}
\end{equation}
where
\begin{equation*}
G(x) = {\sum_{j:j \neq i}} {d_j} F(x-{a_j}).
\end{equation*}
Since ${j^k}(H) = {\J_{km}} {\Phi_{x_m}}$, we have
\begin{equation}
{\J_{jk}} {j^k}(H) {j^l}(H) = - {\J_{ln}} {\Phi_{x_n}}{\Phi_{x_j}}.
\label{D}
\end{equation}
Fix any number $\rho$ so small that ${D^2} \eta = 0$ on ${B_\rho}({a_i})$. 
We have
\begin{equation*}
{\int\limits_{\T^2}} {\eta_{{x_j}{x_l}}} {\J_{jk}} {j^k}(H) {j^l}(H) dx
= - {\int\limits_{{\T^2}\setminus{{B_\rho}({a_i})}}} {\eta_{{x_j}{x_l}}}
{\J_{ln}} {\Phi_{x_n}}{\Phi_{x_j}} 
\end{equation*}
\begin{equation*}
= {\int\limits_{{\T^2}\setminus{{B_\rho}}}} {\eta_{x_l}}
{\J_{ln}} {\Phi_{{x_n}{x_j}}} {\Phi_{x_j}} dx + {\int\limits_{{\partial 
{B_\rho}}}}
{\eta_{x_l}} {\J_{ln}} {\Phi_{x_n}} {\Phi_{x_j}} {\nu^j} d{H^1}
\label{E}
\end{equation*}
where $\nu = ({\nu^1},{\nu^2})$ is the outward unit normal to 
$\partial {B_\rho}$. We recognize ${\Phi_{{x_n}{x_j}}} {\Phi_{x_j}} = \half
{{({\Phi_{x_j}}{\Phi_{x_j}})}_{x_n}}$ and integrate by parts again to find
\begin{eqnarray*}
= - {\int\limits_{{\T^2}\setminus{{B_\rho}}}}{\eta_{{x_l}{x_n}}}
{\J_{ln}} \half {\Phi_{x_j}}{\Phi_{x_j}} dx &-&
 {\int\limits_{{\partial 
{B_\rho}}}} {\eta_{x_l}} {\J_{ln}} \half 
{\Phi_{x_j}}{\Phi_{x_j}} {\nu^n} d{H^1} \\
\label{F1}
& + &{\int\limits_{{\partial 
{B_\rho}}}}
{\eta_{x_l}} {\J_{ln}} {\Phi_{x_n}} {\Phi_{x_j}} {\nu^j} d{H^1}.
\label{F2}
\end{eqnarray*}
Since ${\eta_{{x_l}{x_n}}}
{\J_{ln}}=0$, the integral over $
{{\T^2}\setminus{{B_\rho}}}$ vanishes and we are left with two boundary
integrals $I_\rho$, ${II}_\rho$.
\newline
2. We calculate the boundary integrals. We begin with
\begin{equation*}
{I_\rho} = - {\int\limits_{{\partial{B_\rho}}}} {\eta_{x_l}} {\J_{ln}}
\half {\Phi_{x_j}}{\Phi_{x_j}} {\nu^n} d{H^1}.
\end{equation*}
By using \eqref{C}, we observe
\begin{equation}
{\Phi_{x_j}} {\Phi_{x_j}} = {F_{x_j}}{F_{x_j}} + 2{d_i} {F_{x_j}} {G_{x_j}}
+ {G_{x_j}}{G_{x_j}}.
\label{G}
\end{equation}
Since $F$ is even, the first term integrates to zero. We exploit the fact
that ${G_{x_j}}$ is nearly constant on ${B_\rho}({a_i})$ to calculate the
contribution to $I_\rho$ arising from the remaining two terms in \eqref{G}.
The cross term contributes
\begin{equation}
- {d_i} {\eta_{x_l}}({a_i}) {G_{x_j}}({a_i}) {\J_{ln}} {\int\limits_{{\partial{B_\rho}}}} {F_{x_j}}{\nu^n} d{H^1} - {d_i} {\eta_{x_l}}({a_i})  {\J_{ln}} {\int\limits_{{\partial{B_\rho}}}} {F_{x_j}}[ {G_{x_j}}(x) - {{G_{x_j}}}({a_i})]
{\nu^n} d{H^1}.
\label{H}
\end{equation}
Since $F \thicksim \log |x-{a_i}|$, $|{F_{x_j}}| \thicksim \frac{1}{\rho} $ on
$\partial{B_\rho}$ and $G$ is $C^1$ on $B_\rho$, the second integral 
contributes $O(\rho)$. The ${G_{x_j}}{G_{x_j}}$ term contributes $O(\rho)$
as well.

Next, we calculate
\begin{equation*}
{{II}_\rho} = {\int\limits_{{\partial{B_\rho}}}} {\eta_{x_l}}
{\J_{ln}} {\Phi_{x_n}}{\Phi_{x_j}} {\nu^j} d{H^1}
\end{equation*}
by expanding using \eqref{C}. The ${F_{x_n}}{F_{x_j}}$ term again 
vanishes by symmetry. The ${G_{x_n}}{G_{x_j}}$ term contributes $O(\rho)$
and the cross terms remain to be estimated. The first cross term gives
\begin{equation}
{d_i} {\eta_{x_l}}({a_i}) {G_{x_j}}({a_i}) {\J_{ln}} {\int\limits_{{\partial{B_\rho}}}} {F_{x_n}} {\nu^j} d{H^1} + O(\rho).
\label{J}
\end{equation}

The second cross term contributes
\begin{equation}
{d_i} {\eta_{x_l}}({a_i}) {G_{x_n}}({a_i}) {\J_{ln}} {\int\limits_{{\partial{B_\rho}}}} {F_{x_j}} {\nu^j} d{H^1} + O(\rho).
\label{K}
\end{equation}
\newline
3. Since ${\int_{{\partial{B_\rho}}}} {F_{x_n}} {\nu^j} d{H^1}
= {\int_{{\partial{B_\rho}}}} {F_{x_j}} {\nu^n} d{H^1}$, the first term in
\eqref{H} and \eqref{J} cancel and the only remaining contribution 
is \eqref{K}. Finally, observe that
\begin{equation}
{\int\limits_{\partial{B_\rho}}} {F_{x_j}} {\nu^j} d{H^1} = 
{\int\limits_{{B_\rho}}} \Delta F dx = 2\pi- 2{\pi^2}{\rho^2},
\label{L}
\end{equation}
using \eqref{Fund}.
Therefore
\begin{equation}
{\int\limits_{\T^2}} {\eta_{{x_j}{x_l}}} {\J_{jk}} {j^k}(H) {j^l}(H) dx
= 2\pi {d_i} {\eta_{x_l}}({a_i}) {\J_{ln}} {G_{x_n}}({a_i}) + O(\rho).
\label{M}
\end{equation}
Since $\rho$ can be taken arbitrarily small, we have proved the lemma.

\end{proof}

\section{Acknowledgments}

Many of the ideas in this paper grow out of the second author's joint
work with H.M. Soner and reflect the influence of numerous conversations
over a period of years. The first author was partially supported by
a grant from the University of Illinois Research Board and by
a NSF Postdoctoral Fellowship. The second author was partially
supported by NSF grant \# DMS 96-00080.

\end{document}